\documentclass{aa}
\begin{document}

\thesaurus{06     % A&A Section 6: Form. struct. and evolut. of stars
              (08.22.1;  % Stars: variables: Cepheids
               08.15.1;  % Stars: oscillations
               08.04.1)}  % Stars: distances

   \title{Theoretical models for classical Cepheids: V. 
Multiwavelength relations}

   \author{F. Caputo\inst{1}, M. Marconi\inst{2}, I. Musella\inst{2}}
%          \and
%          C. Ptolemy\inst{2}\fnmsep\thanks{Just to show the usage
%          of the elements in the author field}

   \offprints{M. Marconi, marcella@na.astro.it}

   \institute{Osservatorio Astronomico di Roma, Via di Frascati 33, 
I-00040 Monteporzio Catone, Italy \and Osservatorio Astronomico di
Capodimonte, Via Moiariello 16, I-80131 Napoli, Italy\\
              email: caputo@coma.mporzio.astro.it, marcella@na.astro.it, ilaria@na.astro.it}

   \date{Received .........; accepted ........}

\titlerunning{Theoretical multiwavelength relations}
\authorrunning{Caputo, Marconi, Musella}
\maketitle

   \begin{abstract}

From a theoretical study based on
nonlinear, nonlocal and time-dependent convective pulsating 
models at varying mass and chemical 
composition, we present the predicted 
Period-Luminosity, Period-Color, Color-Color and 
Period-Luminosity-Color relations in the $BVRIJK$ bands for Classical 
Cepheids.  All the theoretical relations are, in various degrees, 
metallicity dependent and the comparison with
observed data for MC and Galactic Cepheids show a 
fair agreement which 
supports the validity of the pulsating models. 

   \end{abstract}

%
%________________________________________________________________

\section{Introduction}

\noindent
Since the Leavitt (1912) discovery of the relationship between period and
apparent magnitude for Cepheids in the Magellanic Clouds, 
these variables have been playing a fundamental role in the 
determination of distances to Local Group galaxies, in the 
calibration of various secondary distance indicators and 
finally in the evaluation of the Hubble constant $H_0$. A considerable
amount of observational studies have been devoted to the 
calibration of their characteristic Period-Luminosity ($PL$) 
and Period-Luminosity-Color ($PLC$) relations,
e.g. via the Baade-Wesselink
method (Ripepi et al. 1997; Gieren et al. 1998
and references therein), or making use of Cepheids 
with distances from Galactic open cluster main-sequence fitting 
(Turner et al. 1998 and references therein), or   
adopting the Large Magellanic Cloud (LMC) distance 
found in other independent ways (see Walker 1999). 

The most important question to be answered is 
whether the {\bf bolometric} $PL$ and $PLC$ are {\it universal}, as early 
suggested by Iben \& Renzini (1984), or depending on the chemical 
abundances of the variables. This is a fundamental point 
since {\bf if there is a metallicity effect on the relations in the 
various photometric bands and} the galaxies whose distances 
we are deriving have metallicities different from the calibrating 
Cepheids, then a  {\bf significant correction could be necessary}. 
In recent times, the occurrence of a metallicity effect has become 
more and more evident, with the observational tests of this effect 
suggesting that metal-rich Cepheids are brighter than 
metal-poor Cepheids at fixed period (see Kochanek 1997; 
Beaulieu et al. 1997;
Sasselov  et al. 1997; Kennicutt et al. 1998). 
As for theoretical estimates,
models based on linear pulsation calculations suggest 
a very low effect (Chiosi et al. 1993; Saio \& Gautschy 1998; 
Alibert et al. 1999), whereas nonlinear, nonlocal and time-dependent 
convective pulsating models suggest that 
both the zero point and the slope of the predicted
$PL$ relations depend on metallicity, with the amplitude of 
the metallicity effect decreasing at the longer wavelengths 
(Bono et al. 1999b [Paper II]).
Moreover, these models show that metal-rich Cepheids are  
on average intrinsically 
fainter than the metal-poor ones, at fixed period.

It is worth noticing that all the observational efforts rely on the 
assumption that the slope of the $PL$ relation for different passbands 
is that found for LMC Cepheids, leading us to suspect that 
forcing the slope of 
the multiwavelength relations to be constant may introduce some 
systematic errors in the attempt of disentangling reddening from 
metallicity effects. On the other hand, at variance with 
the linear-nonadiabatic approach
which supplies  only the blue edge of 
the instability strip \footnote{Linear red edge 
estimates by Chiosi et al. (1993) 
and by Alibert et al. (1999) were  fixed, more or
less tentatively, at the effective temperature where the growth
rate attains its maximum value.  However, the 
nonlinear results suggest that 
such an assumption is far from being adequate 
(Bono et al. 1999c [Paper III]).}, 
the models presented in Paper II 
provide fine constraints 
on both blue and red limits of the pulsation, thus avoiding dangerous 
{\it ad hoc} assumptions on the width of the instability strip
(see also Tanvir 1999). These models    
give also the pulsation amplitude and the predicted mean magnitudes 
of the pulsator. 

The theoretical $PL$ and $PLC$ relations presented in Paper II 
deal with the intensity-weighted $<M_B>$, $<M_V>$ and $<M_K>$ 
magnitudes. In order to have a full insight into 
the metallicity effect, and considering that the HST observations 
of extragalactic Cepheids
consider the $I$ band, in this paper we
present the full set of theoretical relations (i.e. Period-Luminosity,
Period-Color, Color-Color and Period-Luminosity-Color) in the $BVRIJK$
bands. The predicted relations are reported in Sect. 2, while Sect. 3
gives brief comparisons of our theoretical results against observed data
for calibrating Cepheids. Some final remarks close the paper.

%__________________________________________________________________

\section{Predicted relations}
The nonlinear convective pulsating models adopted in this paper are computed
with four values of the stellar mass ($M/M_{\odot}=$5, 7, 9, 11) and three
chemical compositions ($Y$=0.25, $Z$=0.004; $Y$=0.25, $Z$=0.008; 
$Y$=0.28, $Z$=0.02),
taken as representative of Cepheids in the Magellanic Clouds and in the
Galaxy. The basic assumptions on the input physics, computing procedures 
and the adopted mass-luminosity relation have been already discussed in 
Bono et al. (1998), Bono et al. 
(1999a [Paper I]) and Paper III
and will not be repeated. {\bf Here we wish only to remark that the
whole instability strip moves toward lower effective temperatures as the
metallicity increases and, as a consequence, that the bolometric magnitude
of the pulsators increases as the metal abundance increases (see Fig. 2
in Paper II).}

\subsection{\it Period-Luminosity}
\noindent
From the bolometric light curves of our pulsating models and adopting the
grid of atmosphere models provided by Castelli et al. (1997a,b),
we first derive the predicted intensity-weighted mean magnitude $<M_{\lambda}>$
of pulsators for the $BVRIJK$ passbands 
\footnote{These 
atmosphere models do not allow the calculation of 
synthetic $H$ magnitudes.}.  The fundamental 
pulsators plotted in Fig. 1, where the 
different lines depict the blue and red limits of the pulsation region 
 at the various compositions, show that 
the metal abundance effects on the predicted location and width of the 
instability strip are significantly dependent on the photometric passband. 
In particular, the models confirm 
early results (e.g., see Laney \& Stobie 1994; Tanvir 1999 
and reference therein) that infrared magnitudes are needed 
to reduce both the intrinsic scatter and the metal abundance 
dependence of the $PL$ relation.

\begin{figure} 
\vspace{1cm}
\caption{Location in the log$P$-$<M_V>$ and log$P$-$<M_K>$ plane 
of fundamental pulsators with different 
chemical compositions and masses (from left to right: 5, 7, 9 and 
11$M_{\odot}$). The lines depict the predicted edges of the instability strip.}
\end{figure}

This matter has been already discussed in Paper II on the basis of the 
predicted $PL_V$ and $PL_K$ relations derived under the assumption of a 
uniformly populated instability strip. However, since the $PL$ is a 
``statistical'' relation which provides the average of the 
Cepheid magnitudes 
$\overline{{M_{\lambda}}}$ at a given period, we decide 
to estimate the effect of a different pulsator population. 
Thus, following the procedure outlined by Kennicutt et al. (1998), 
for each given chemical composition we populate the instability 
strip\footnote{The location of the instability strip is 
constrained by the predicted blue and red edges.} 
with 1000 pulsators and a mass distribution as given by 
$dn/dm=m^{-3}$ over the mass range 5-11$M_{\odot}$. 
The corresponding luminosities are obtained from the 
mass-luminosity relation derived from
canonical evolutionary models, i.e. with vanishing efficiency of convective
core overshooting (see Bono \& Marconi 1997; Paper II; Paper III).

\begin{figure}
\vspace{1cm}
\caption{Period-magnitude distribution 
of fundamental pulsators at varying metallicity and photometric bandpass. 
Dashed lines and solid lines refer to the quadratic and linear 
$PL$ relations given in Table 1 and Table 2, 
respectively.}
\end{figure}

Fig. 2 shows the resulting $\log{P}-M_{\lambda}$ 
distribution of fundamental pulsators with the three selected 
metallicities. We derive that the pulsator distribution 
becomes more and more linear by going toward the infrared and that, 
aiming at  reducing the intrinsic scatter of $PL$, 
the pulsator distributions in the $\log{P}-M_B$, $\log{P}-M_V$ 
and $\log{P}-M_R$ planes are much better represented by a 
quadratic relation. The dashed lines in Fig. 2 show the quadratic 
least square fit ($\overline{M_{\lambda}}=a+b\log{P}+c\log{P}^2$), 
while the solid lines refer to the linear approximation 
($\overline{M_{\lambda}}=a+b\log{P}$). The values for the 
coefficients $a$, $b$ and $c$ of each $PL_{\lambda}$ relation are listed 
in Table 1 (quadratic solution) and Table 2 (linear solution), together 
with the rms dispersion ($\sigma$) of $M_{\lambda}$ about the fit. 
However, we wish to remark that the present solutions refer to a  
specific pulsator distribution and that different populations 
may modify the results. As a matter of example, if 
the longer periods (log$P\ge$1.5) are rejected in the final fit, 
then the predicted linear $PL$ relations become steeper and the 
intrinsic dispersion in the $BVR$ bands 
is reduced (see Table 3).     
 
As a whole, each predicted $PL_{\lambda}$ relation seems to become steeper 
at the lower metal abundances, with the amplitude of the effect 
decreasing from $B$ to $K$ filter. Moreover, the slope and the intrinsic 
dispersion of the predicted $PL$ relation at a given $Z$ decrease 
as the filter wavelength increases (see Fig. 3 for the linear relations), 
in close agreement with the observed trend 
(e.g. see Fig. 6 in Madore \& Freedman 1991).

\begin{figure}
\vspace{1cm}
\caption{Slope and intrinsic dispersion of the predicted 
linear $PL$ relations with 
different bandpass.}
\end{figure}

In closing this section, let us observe that the adopted way to 
populate the instability strip allows $PL$ relations only for the 
``static''  magnitudes (the value the star would have were it not 
pulsating), while the observations deal with the mean magnitudes 
averaged over the pulsation cycle [($m_{\lambda}$) 
if magnitude-weighted and $<m_{\lambda}>$ if intensity-weighted]. 
It has been recently shown (Caputo et al. 1999 [Paper IV]) 
that the differences among static and mean magnitudes and 
between magnitude-weighted and intensity-weighted averages 
are always smaller than the intrinsic scatter of the $PL$ 
relation ($\sim0.1$ mag for V magnitudes and $\le{0.01}$ mag for 
K magnitudes). Here, given the marginal agreement of the 
coefficients in Table 1, Table 2 and Table 3 with the values given 
in Paper IV, we conclude that the additional effect on 
the intrinsic scatter of the $PL$ relations, as 
due to different Cepheid populations, is negligible, 
provided that a statistically significant sample of 
variables is taken into account.

\subsection{\it Period-Color and Color-Color}
\noindent
The three methods of deriving the mean 
color over the pulsation cycle use either ($m_j$-$m_i$), the average over 
the color curve taken in magnitude units, or $<m_j-m_i>$, the mean 
intensity over the color curve transformed into magnitude, 
or $<m_j>$-$<m_i>$, the difference of the mean intensities
transformed into magnitude, performed separately over the 
two bands. In Paper IV it has been shown that there are some 
significant differences between static and synthetic mean colors, 
and that the predicted  
($m_j$-$m_i$) colors are generally redder than $<m_j>$-$<m_i>$ colors, 
with the difference depending on the shape of the light curves, in 
close agreement with observed colors for 
Galactic Cepheids. As a matter of example, 
the predicted difference $(B-V)-[<B>-<V>]$ ranges from 0.02 mag 
to 0.08 mag, whereas $(V-K)-[<V>-<K>]$ 
is in the range of 0.014 mag to 0.060 mag.

 Fig. 4 shows the pulsator synthetic mean colors as 
a function of the period for the three different metallicities. 
Since, given the finite width of the instability strip,    
also the $PC$ relation is a ``statistical''  relation 
between period and the average of the mean color indices $CI$, we 
follow the procedure of Sect. 2.1. It is evident from Fig. 4
that the pulsator distribution shows a quadratic behavior, as well as that the 
intrinsic dispersion {\bf of $PC$ relations allows reddening estimates 
within $\sim{0.07}$ mag, on average}. However, as far as they could help, we 
present the predicted relations $\overline{CI}=A+B$log$P$ 
at the various compositions. 
The relations are 
plotted in Fig. 4 and the   
$A$ and $B$ coefficients are listed in Table 4 together with the rms 
deviation of the color about the fit ($\sigma$). 
Note that also the slope of the predicted 
$PC$ relations depends on the metal abundance, increasing
with increasing $Z$.

\begin{figure}
\vspace{1cm}
\caption{Location in the period-color plane 
of fundamental pulsators with different
chemical compositions and masses. The lines depict the linear 
$PC$ relations given in Table 4.}
\end{figure}

\begin{figure}
\vspace{1cm}
\caption{The color-color plane of fundamental pulsators. The lines depict 
the $CC$ relations at varying metallicities (see Table 5).}
\end{figure}

 As for the color-color ($CC$) relations, Fig. 5 shows  
that the spurious effect due to 
the finite width of the instability strip is 
almost removed. For this reason, 
from our pulsating models we derive 
a set of color-color ($CC$) relations correlating 
$<B>$-$<V>$ with $<V>$-$<R>$, $<V>$-$<I>$, $<V>$-$<J>$ and $<V>$-$<K>$, 
but similar relations adopting 
($m_j$-$m_i$) or $<m_j-m_i>$ colors can be obtained upon request. 
The predicted $CC$ linear relations are given in Table 5, 
while Fig. 5 shows the pulsator distribution in the color-color plane. 
One may notice that the intrinsic scatter of the theoretical relations 
is very small, with the rms dispersion of $<B>$-$<V>$ about the fit 
smaller than 0.01 mag.

\subsection{\it Period-Luminosity-Color}
\noindent
Since the pioneering paper by Sandage (1958), Sandage \& Gratton (1963) 
and Sandage \& Tammann (1968), it is well known that if the 
Cepheid magnitude is given as a function of
the pulsator period and color, i.e. if the Period-Luminosity-Color
relation is considered,  then the tight  
correlation among the parameters of individual Cepheids is 
reproduced (see also Laney \& Stobie 1986; 
Madore \& Freedman 1991; Feast 1995 and references therein).

In Paper II it has been shown that the intrinsic scatter of the 
$PLC$ relations in the visual ($<M_V>$, $<B>$-$<V>$) and 
infrared ($<M_V>$, $<V>$-$<K>$) is $\sim0.04$ mag. 
Furthermore, we showed in Paper IV that the systematic effect on 
the predicted visual $PLC$ relations, as due to the adopted method of 
averaging magnitudes and colors over the pulsation cycle,  
are always larger than the intrinsic scatter of the relation. 
For this reason in this paper we present only the predicted $PLC$ 
relations $<M_V>=\alpha+\beta\log{P}+\gamma[<M_V>-<M_{\lambda}>]$, 
but similar relations for magnitude-weighted values can 
be obtained upon request. 

\begin{figure}
\vspace{1cm}
\caption{Projection onto a plane of the 
predicted $PLC$ relations ($<B>-<V>$ and $<V>-<R>$ colors) 
for fundamental pulsators with different metallicities 
(see Table 6).}
\end{figure}

\begin{figure}
\vspace{1cm}
\caption{As in Fig. 6, but with 
$<V>-<I>$ and $<V>-<J>$ colors.}
\end{figure}

\begin{figure}
\vspace{1cm}
\caption{As in Fig. 6, but with
$<V>-<K>$ colors.}
\end{figure}

From the least square solutions through the fundamental models 
we derive the coefficients $\alpha$, $\beta$ and $\gamma$ 
presented in Table 6, together with the residual dispersion 
($\sigma$) of $<M_V>$ about the fit. 
Figs. 6-8 illustrate the remarkably 
small scatter of the $PLC$ relations (see Fig. 1 for comparison). 
Moreover, one may notice that adopting $B-V$ color, the 
metal-rich Cepheids are brighter than metal-poor ones with the 
same period and color (see lower panel of Fig. 6), 
whereas the opposite trend holds with $V-K$ color (see Fig. 8). 
As a ``natural''  consequence, the predicted relationship 
with $V-I$ color (lower panel of Fig. 7) 
turns out to be almost independent of the metal abundance.

In order to complete the theoretical framework for classical Cepheids, 
we have finally considered the Wesenheit quantities $W$ (Madore 1982) 
which are often used to get a reddening-free formulation of the $PL$ 
relation. With $A_{\lambda}$ 
giving the absorption in the $\lambda$-passband, one has

$$W(B,V)=V-[A_V/(A_B-A_V)](B-V)$$
$$W(B,V)=V_0+A_V-[A_V/(A_B-A_V)][(B-V)_0+(A_B-A_V)]$$
$$W(B,V)=V_0-[A_V/(A_B-A_V)](B-V)_0,$$

$$W(V,R)=R-[A_R/(A_V-A_R)](V-R)$$
$$W(V,R)=R_0+A_R-[A_R/(A_V-A_R)][(V-R)_0+(A_V-A_R)]$$
$$W(V,R)=R_0-[A_R/(A_V-A_R)](V-R)_0,$$

               $$etc....$$

\noindent
Table 7 gives the coefficients of the theoretical reddening-free 
$PL$ [hereinafter $WPL$] relations derived from our fundamental models. 
Note that present results adopt $A_V/E_{B-V}=3.1$, $A_R/E_{V-R}=5.29$, 
$A_I/E_{V-I}=1.54$, $A_J/E_{V-J}=0.39$ and $A_K/E_{V-K}=0.13$
from the Cardelli et al. (1989) extinction model,
but formulations using different ratios of total to selective 
absorption can be obtained upon request.

\begin{figure}
\vspace{1cm}
\caption{Predicted $WPL$ relations  
of fundamental pulsators with different
chemical compositions (see Table 7).}
\end{figure}

\begin{figure}
\vspace{1cm}
\caption{Predicted $WPL$ relations
of fundamental pulsators with different
chemical compositions (see Table 7).}
\end{figure}

Figs. 9-10 show the theoretical Wesenheit quantities as a function 
of the period, together with the predicted relations. 
From a comparison of Table 7 with Table 2 one derives that 
 the intrinsic scatter of $W(B,V)PL$ and $W(V,R)PL$ 
is significantly lower than the dispersion of $PL$, while no significant 
improvement occurs with $W(V,I)PL$ and somehow larger dispersions 
are found $W(V,J)PL$ and $W(V,K)PL$.

Moreover, the comparison between Figs. 9-10
and Figs. 6-8 discloses the deep difference between 
$PLC$ and $WPL$ relations (see also Madore \& Freedman 1991). 
The former ones are able to define accurately the properties of 
individual Cepheids within the instability strip, whereas 
the latter ones are thought to cancel the reddening effect. 
As a consequence, provided that the variables are at the same distance 
and have the same metal abundance, the scatter in observed $PLC$ relations 
should depend on {\bf errors in the adopted}  reddening, whereas the scatter in observed 
$WPL$ relations is a residual effect
of the finite width of the strip.

\section{Comparison with Galactic Cepheids}
\noindent
The principal aim of this paper is to present a wide homogeneus theoretical
scenario for Cepheid studies. The analysis of observed pulsational
properties is out of our present intentions, but it seems necessary 
to verify the reliability of our models against well studied variables. 
For this purpose, we take into consideration 
the sample of calibrating Galactic Cepheids studied by  
 Gieren et al. (1998, [GFG]) and 
by  Laney \& Stobie (1994, [LS]). To the 
intensity mean magnitudes $BVJK$ magnitudes we add 
the $<I>$ magnitudes as derived 
from $<V>-<I>-(V-I)$=-0.03 mag, where the magnitude-weighted (V-I)
colors are by Caldwell \& Coulson (1987). Such a constant 
correction, suggested by GFG, 
is confirmed (see Fig. 11) from our synthetic mean colors. 
 Note that, according to  GFG, we excluded  
SV Vul, GY Sge and S Vul for problems 
with a variable period and EV Sct, SZ Tau and QZ Nor for uncertainty 
with regard to the 
pulsation mode.

 Before analyzing the calibrating 
Galactic Cepheids, let us briefly consider 
the recent results on Magellanic Clouds Cepheids provided by the 
OGLE II microlensing survey (Udalski et al. 1999). The final $PL_V$ 
and $PL_I$ relations given by these authors for Cepheids with
 $\log{P}\le{1.5}$, are $M_V$=-1.18-2.765log$P$ and
$M_I$=-1.66-2.963log$P$, with a LMC distance modulus
$\mu_{LMC}$=18.22 mag. The data in Table 3 
with $Z$=0.008 yield $M_V$=-1.37-2.75log$P$ and
$M_I$=-1.95-2.98log$P$, 
namely a predicted slope in close  
agreement with OGLE measurements but a zero-point which is 
brighter by roughly 0.24 mag, suggesting $\mu_{LMC}$=18.46 mag, 
a value which fits the recent upward revision of the LMC distance
derived from field red clump stars 
(Zaritsky 1999). 
    
As for the $W(V,I)PL$ relation presented by Udalski et al. (1999),  
we find that its slope (-3.28) is in fair agreement with the 
results (-3.17) of our models with $Z$=0.008 
(see also the lower panel 
of Fig. 12). 
Conversely, the OGLE $PLC_I$ relation discloses a significative 
difference with respect to our predicted relations. As 
shown in the upper panel 
of Fig. 12, it seems that the OGLE color-term is 
not able to fully remove the effect of the finite width of 
the instability strip 
(as a matter of comparison, see the lower panel of Fig. 7). 
At the moment we cannot explain such a disagreement and we 
expect the release of LMC and SMC data to test the full set of 
our predictions.

\begin{figure}
\vspace{1cm}
\caption{Predicted difference between $<V>-<I>$ and $(V-I)$ colors for 
fundamental pulsators with different masses and chemical compositions. 
The solid line refers to a difference of -0.03 mag.}
\end{figure}

\begin{figure}
\vspace{1cm}
\caption{{(\it lower panel)} Wesenheit quantities of fundamental 
pulsators in comparison with 
the OGLE $WPL$ relation. {(\it upper panel)} OGLE $PLC$ relation applied 
to fundamental pulsators.}
\end{figure}

\begin{figure}
\vspace{1cm}
\caption{Calibrating Galactic Cepheids (dots) in comparison with predicted 
$PL$ relations with $Z$=0.02 (dotted line). 
Data from Laney \& Stobie (1994 [LS]).}
\end{figure}

\begin{figure}
\vspace{1cm}
\caption{Calibrating Galactic Cepheids (dots) in comparison with predicted
$PL$ relations with $Z$=0.02 (dotted line). 
Data from Gieren et al. (1998 [GFG]).}
\end{figure}

\begin{figure}
\vspace{1cm}
\caption{Calibrating Galactic Cepheids (dots) from Laney \& 
Stobie (1994 [LS]) and Gieren et al. (1998 [GFG]), 
in comparison with OGLE 
$PL$ relations for LMC and SMC Cepheids (dashed-dotted line).}
\end{figure}

\begin{figure}
\vspace{1cm}
\caption{As in Fig. 13, but for predicted $PLC$ relations.}
\end{figure}

\begin{figure}
\vspace{1cm}
\caption{As in Fig. 14, but for predicted $PLC$ relations.}
\end{figure}

Passing to the Galactic Cepheids, Fig. 13 and Fig. 14 show the predicted 
linear $PL$ relations with $Z$=0.02 (dotted line) in 
comparison with the calibrating Cepheids (dots) from LS and GFG, respectively. 
Note that both the reddening and true distance modulus given by the authors 
are adopted. We find a fair agreement in the infrared, whereas in the 
$V$ and $I$ bands the observed 
Cepheids with the longer periods appear brighter than the predicted 
relations. This could suggest that these luminous variables have lower 
metallicities than the currently adopted value $Z$=0.02 (see 
Fry \& Carney 1997) or be near the blue edge of the instability strip. 
 One could also suspect that the predicted slopes with 
$Z$=0.02 are smaller 
than the actual ones. This seems supported by Fig. 15 
where the steeper $PL_V$ relation provided by  
Udalski et al. (1999) is taken into account.

However, in order to remove the effects of the finite 
width of the instability strip, let us apply our 
predicted $PLC$ relations to the calibrating 
Cepheids. 
Figs. 16-17 
disclose that predictions and 
observed data are now in a better agreement. Meanwhile, 
there are some evidence that 
the calibrating Galactic Cepheids lie on the average  
below the theoretical $PLC$ relations 
with $Z$=0.02. In other words, 
our predicted relations  yield 
somehow larger true distance moduli than those given by LS and GFG. 
 Specifically, the left panels of   
Fig. 18 show that the difference between LS and GFG data and the results 
of our predicted $PLC_{VI}$ relation with $Z$=0.02, by adopting the 
reddening values listed by these authors, have a significant 
period dependence. This trend cannot be ascribed to our color and 
period term since a quite similar behavior is present in the right 
panels which refer to 
the $PLC_{VI}$ relation given by Udalski et al. (1999). {\bf Notwithstanding 
the different values of the period and color terms, one finds a similar trend,
with
the OGLE relation suggesting smaller distances than those listed 
by LS and GFG.}

\begin{figure}
\vspace{1cm}
\caption{True distance modulus of Galactic Cepheids from the   
$PLC_{VI}$ relation in comparison with Laney \& Stobie (1994 [LS] 
and Gieren et al. (1998 [GFG]) data. The left panel refer 
to the predicted relation with $Z$=0.02, while the right panels refer 
to the OGLE result for LMC and SMC Cepheids.}
\end{figure}

\begin{figure}
\vspace{1cm}
\caption{Galactic Cepheids 
$BVI$ reddenings from predicted $CC$ relations with $Z$=0.02 
in comparison with LS93 (circles) and LS94 (dots) values.}
\end{figure}

\begin{figure}
\vspace{1cm}
\caption{Predicted $(B-V)-(V-I)$ relation with $Z$=0.02 
(dotted line) in comparison with the intrinsic locus of 
Galactic Cepheids. The solid line and the dashed line refer 
to the labelled solutions, as given by Dean, Warren \& Cousins 
(1978).}
\end{figure}

{\bf On the other hand}, we show in Fig. 19 that 
the $BVI$ reddenings \footnote{We adopt the ratios of total to selective
absorption in the $VI$ bands from Caldwell \& Coulson (1987) and Laney 
\& Stobie (1993).} 
derived from the predicted color-color relation with $Z$=0.02 are slightly 
lower than those given by LS and GFG. The origin of 
the discrepancy could be found in Fig. 20, where 
our predicted color-color relation with $Z$=0.02 (dotted line) is plotted 
together with the two quadratic relations given 
by Dean et al. (1978) for the intrinsic color-color 
locus of the Galactic Cepheids (solid line and dashed line). Since 
observational studies generally adopt the solution depicted with the 
solid line,  the reason for which our predicted $CC$ relation 
yields slightly smaller $BVI$ reddenings is explained. This {\bf could provide}
the key to understand the difference in the true distance moduli. 

\section{Summary}
\noindent
From nonlinear, nonlocal and time-dependent convective pulsating models
we have derived a set of homogeneus 
$PL$, $PC$, $CC$, $PLC$ and $WPL$ relations in the 
$BVRIJK$ passbands at varying chemical composition. 
We find that the predicted relations are, 
in various degrees, metallicity
dependent, suggesting that the adoption of {\it universal} relations  
for Cepheid studies should
be treated with caution. 

The slope of the predicted relations with $Z$=0.004-0.008 matches  
the results of LMC and SMC Cepheids collected during 
the OGLE II microlensing survey (Udalski et 
al. 1999), but our zero points suggest a larger LMC distance, in agreement 
with the suggestions by Zaritsky (1999). 

As for the comparison of the predicted relations at $Z$=0.02 with 
the observed data of calibrating Galactic Cepheids, we find a 
fair agreement even though the slope of our $PL_V$ and 
$PL_I$ relations seems shallower than observed. This is the consequence 
of the significant metallicity dependence 
in the predicted relations and we believe that a careful analysis 
(of larger samples of Galactic Cepheids) which 
takes into account the actual metallicity dispersion is required to 
settle the question.

Special requests for other theoretical relations can be addressed 
to M. Marconi.

\begin{acknowledgements}
We deeply thank the referee (Dr. Bersier) fore several valuable 
comments which improved the first version of the paper. 
Financial support for this work was provided by the Ministero dell'Universit\`a e della Ricerca Scientifica e Tecnologica (MURST) under the scientific project
``Stellar Evolution'' (Vittorio Castellani, coordinator).
\end{acknowledgements}

\begin{table}
\caption{Theoretical PL relations for fundamental pulsators. 
Quadratic solutions: $\overline{M_{\lambda}}$=$a$+$b$log$P$+$c$log$P^2$.}

\vspace{0.5truecm}

{\centering %\hspace{-4truecm}
\begin{tabular}{ccccc}
\hline
\hline
Z & $a$ & $b$  & $c$ & $\sigma$ \\
\hline
&&&&\\
\multicolumn{5}{c}{$\overline{M_B}$}\\
&&&&\\
0.004&-0.01$\pm$0.06&-4.81$\pm$0.10&1.14$\pm$0.06&0.24\\
0.008&-0.21$\pm$0.06&-4.17$\pm$0.13&0.94$\pm$0.06&0.26\\
0.002&-0.93$\pm$0.04&-2.43$\pm$0.09&0.39$\pm$0.04&0.21\\
&&&&\\
\multicolumn{5}{c}{$\overline{M_V}$}\\
&&&&\\
0.004&-0.69$\pm$0.04&-4.43$\pm$0.10&0.81$\pm$0.05&0.18\\
0.008&-0.86$\pm$0.04&-3.98$\pm$0.09&0.67$\pm$0.05&0.19\\
0.02&-1.41$\pm$0.03&-2.75$\pm$0.07&0.30$\pm$0.03&0.16\\
&&&&\\
\multicolumn{5}{c}{$\overline{M_R}$}\\
&&&&\\
0.004&-1.08$\pm$0.04&-4.28$\pm$0.08&0.67$\pm$0.04&0.15\\
0.008&-1.22$\pm$0.04&-3.91$\pm$0.08&0.56$\pm$0.04&0.16\\
0.02&-1.69$\pm$0.03&-2.88$\pm$0.06&0.26$\pm$0.03&0.13\\
&&&&\\
\multicolumn{5}{c}{$\overline{M_I}$}\\
&&&&\\
0.004&-1.48$\pm$0.03&-4.16$\pm$0.07&0.56$\pm$0.03&0.13\\
0.008&-1.59$\pm$0.03&-3.84$\pm$0.07&0.47$\pm$0.03&0.13\\
0.02&-1.98$\pm$0.02&-2.99$\pm$0.05&0.23$\pm$0.02&0.11\\
&&&&\\
\multicolumn{5}{c}{$\overline{M_J}$}\\
&&&&\\
0.004&-1.97$\pm$0.02&-3.97$\pm$0.05&0.40$\pm$0.02&0.09\\
0.008&-2.04$\pm$0.02&-3.73$\pm$0.05&0.33$\pm$0.02&0.10\\
0.02&-2.32$\pm$0.01&-3.08$\pm$0.03&0.12$\pm$0.02&0.08\\
&&&&\\
\multicolumn{5}{c}{$\overline{M_K}$}\\
&&&&\\
0.004&-2.42$\pm$0.01&-3.81$\pm$0.03&0.26$\pm$0.01&0.06\\
0.008&-2.45$\pm$0.01&-3.65$\pm$0.03&0.21$\pm$0.01&0.06\\
0.02&-2.63$\pm$0.01&-3.19$\pm$0.02&0.06$\pm$0.01&0.05\\
&&&&\\
\hline\hline
\end{tabular}\par}
\end{table}
\vspace{0.30cm}

\begin{table}
\begin{center}
\caption{Theoretical PL relations for fundamental pulsators. Linear solutions: $\overline{M_{\lambda}}$=$a$+$b$log$P$.}

\vspace{0.5truecm}

\begin{tabular}{cccc}
\hline
\hline
Z & $a$ & $b$  & $\sigma$ \\
\hline
&&&\\
\multicolumn{4}{c}{$\overline{M_B}$}\\
&&&\\
0.004&-1.07$\pm$0.02&-2.49$\pm$0.02&0.27\\
0.008&-1.08$\pm$0.02&-2.24$\pm$0.02&0.28\\
0.02&-1.28$\pm$0.02&-1.64$\pm$0.02&0.22\\
&&&\\
\multicolumn{4}{c}{$\overline{M_V}$}\\
&&&\\
0.004&-1.44$\pm$0.02&-2.79$\pm$0.02&0.20\\
0.008&-1.48$\pm$0.02&-2.60$\pm$0.02&0.20\\
0.02&-1.68$\pm$0.01&-2.15$\pm$0.01&0.16\\
&&&\\
\multicolumn{4}{c}{$\overline{M_R}$}\\
&&&\\
0.004&-1.71$\pm$0.01&-2.90$\pm$0.01&0.17\\
0.008&-1.74$\pm$0.01&-2.75$\pm$0.01&0.17\\
0.02&-1.92$\pm$0.01&-2.36$\pm$0.01&0.14\\
&&&\\
\multicolumn{4}{c}{$\overline{M_I}$}\\
&&&\\
0.004&-2.00$\pm$0.01&-3.00$\pm$0.01&0.14\\
0.008&-2.03$\pm$0.01&-2.88$\pm$0.01&0.14\\
0.02&-2.18$\pm$0.01&-2.53$\pm$0.01&0.12\\
&&&\\
\multicolumn{4}{c}{$\overline{M_J}$}\\
&&&\\
0.004&-2.34$\pm$0.01&-3.15$\pm$0.01&0.10\\
0.008&-2.35$\pm$0.01&-3.06$\pm$0.01&0.10\\
0.02&-2.43$\pm$0.01&-2.84$\pm$0.01&0.08\\
&&&\\
\multicolumn{4}{c}{$\overline{M_K}$}\\
&&&\\
0.004&-2.66$\pm$0.01&-3.29$\pm$0.01&0.07\\
0.008&-2.65$\pm$0.01&-3.23$\pm$0.01&0.07\\
0.02&-2.68$\pm$0.01&-3.08$\pm$0.01&0.05\\
&&&\\
\hline
\hline
\end{tabular}\par
\end{center}
\end{table}

\begin{table}
\begin{center}
\caption{Theoretical PL relations for fundamental pulsators with $\log{P}< 1.5$. Linear solutions: $\overline{M_{\lambda}}$=$a$+$b$log$P$.}

\vspace{0.5truecm}

\begin{tabular}{cccc}
\hline
\hline
Z & $a$ & $b$  & $\sigma$ \\
\hline
&&&\\
\multicolumn{4}{c}{$\overline{M_B}$}\\
&&&\\
0.004&-0.90$\pm$0.03&-2.71$\pm$0.02&0.24\\
0.008&-0.93$\pm$0.03&-2.44$\pm$0.02&0.25\\
0.02&-1.21$\pm$0.02&-1.73$\pm$0.02&0.19\\
&&&\\
\multicolumn{4}{c}{$\overline{M_V}$}\\
&&&\\
0.004&-1.32$\pm$0.02&-2.94$\pm$0.02&0.17\\
0.008&-1.37$\pm$0.02&-2.75$\pm$0.02&0.18\\
0.02&-1.62$\pm$0.01&-2.22$\pm$0.01&0.14\\
&&&\\
\multicolumn{4}{c}{$\overline{M_R}$}\\
&&&\\
0.004&-1.61$\pm$0.02&-3.03$\pm$0.01&0.15\\
0.008&-1.65$\pm$0.02&-2.87$\pm$0.01&0.16\\
0.02&-1.88$\pm$0.01&-2.42$\pm$0.01&0.12\\
&&&\\
\multicolumn{4}{c}{$\overline{M_I}$}\\
&&&\\
0.004&-1.92$\pm$0.01&-3.11$\pm$0.01&0.12\\
0.008&-1.95$\pm$0.01&-2.98$\pm$0.01&0.13\\
0.02&-2.14$\pm$0.01&-2.58$\pm$0.01&0.10\\
&&&\\
\multicolumn{4}{c}{$\overline{M_J}$}\\
&&&\\
0.004&-2.28$\pm$0.01&-3.23$\pm$0.01&0.09\\
0.008&-2.29$\pm$0.01&-3.13$\pm$0.01&0.10\\
0.02&-2.41$\pm$0.01&-2.87$\pm$0.01&0.07\\
&&&\\
\multicolumn{4}{c}{$\overline{M_K}$}\\
&&&\\
0.004&-2.61$\pm$0.01&-3.33$\pm$0.01&0.06\\
0.008&-2.61$\pm$0.01&-3.27$\pm$0.01&0.06\\
0.02&-2.67$\pm$0.01&-3.09$\pm$0.01&0.04\\
&&&\\
\hline
\hline
\end{tabular}\par
\end{center}
\end{table}

\begin{table}
\caption{Theoretical Period-Color relations.}
\begin{center}
\begin{tabular}{cccc}
\hline
\hline
Z & $A$ & $B$  & $\sigma$ \\
%\tablehead{
%\colhead{Z\tablenotemark{a}} &
%\colhead{$A$\tablenotemark{b}} &
%\colhead{$B$\tablenotemark{c}} &
%\colhead{$\sigma$\tablenotemark{d}} }
%\startdata
\hline
               &                &         &         \\
  \multicolumn{4}{c}{$\overline{B-V}$=$A$+$B$log$P$} \\
               &                &         &         \\
0.004 &0.37 $\pm$0.01& 0.30 $\pm$0.01& 0.07  \\
0.008 &0.40 $\pm$0.01& 0.37 $\pm$0.01& 0.07\\
0.02  &0.40 $\pm$0.01& 0.51 $\pm$0.01& 0.05\\
                &         &         &         \\
  \multicolumn{4}{c}{$\overline{V-R}$=$A$+$B$log$P$} \\
               &                &         &         \\
0.004 &0.26 $\pm$0.01& 0.12 $\pm$0.01& 0.03 \\
0.008 &0.26 $\pm$0.01& 0.15 $\pm$0.01& 0.03\\
0.02  &0.24 $\pm$0.01& 0.21 $\pm$0.01& 0.02\\
                &         &         &         \\
  \multicolumn{4}{c}{$\overline{V-I}$=$A$+$B$log$P$} \\
               &                &         &         \\
0.004 &0.56 $\pm$0.01& 0.22 $\pm$0.01& 0.06 \\
0.008 &0.55 $\pm$0.01& 0.28 $\pm$0.01& 0.06 \\
0.02  &0.50 $\pm$0.01& 0.38 $\pm$0.01& 0.04 \\
                &         &         &         \\
  \multicolumn{4}{c}{$\overline{V-J}$=$A$+$B$log$P$} \\
               &                &         &         \\
0.004 &0.90 $\pm$0.01& 0.36 $\pm$0.01& 0.09\\
0.008 &0.87 $\pm$0.01& 0.46 $\pm$0.01& 0.10\\
0.02  &0.75 $\pm$0.01& 0.69 $\pm$0.01& 0.08\\
                &         &         &         \\
  \multicolumn{4}{c}{$\overline{V-K}$=$A$+$B$log$P$} \\
               &                &         &         \\
0.004 &1.21 $\pm$0.01& 0.50 $\pm$0.01&  0.13\\
0.008 &1.17 $\pm$0.01& 0.62 $\pm$0.01& 0.14\\
0.02  &1.00 $\pm$0.01& 0.93 $\pm$0.01& 0.12\\
\hline
\hline
\end{tabular}
\end{center}
%\end{flushleft}
\end{table}

\begin{table}
\caption{Theoretical Color-Color relations.}
\begin{center}
\begin{tabular}{cccc}
\hline
\hline
Z & $C$ & $D$  & $\sigma$ \\
\hline
               &                &         &         \\
  \multicolumn{4}{c}{${<B>-<V>}$=$C$+$D{[{<V>-<R>}]}$} \\
               &                &         &         \\
0.004 &-0.25 $\pm$0.01& 2.38 $\pm$0.03&  0.01\\
0.008 &-0.22 $\pm$0.01& 2.39 $\pm$0.04& 0.01\\
0.02  &-0.16 $\pm$0.01& 2.31 $\pm$0.02& 0.01\\
               &                &         &         \\
  \multicolumn{4}{c}{${<B>-<V>}$=$C$+$D{[{<V>-<I>}]}$} \\
               &                &         &         \\
0.004 &-0.33 $\pm$0.01& 1.28 $\pm$0.02&  0.01\\
0.008 &-0.30 $\pm$0.01& 1.28 $\pm$0.03& 0.01\\
0.02  &-0.24 $\pm$0.01& 1.28 $\pm$0.02& 0.01\\
               &                &         &         \\
  \multicolumn{4}{c}{${<B>-<V>}$=$C$+$D{[{<V>-<J>}]}$} \\
               &                &         &         \\
0.004 &-0.29 $\pm$0.01& 0.75 $\pm$0.02&  0.01\\
0.008 &-0.23 $\pm$0.02& 0.74 $\pm$0.03& 0.01\\
0.02  &-0.13 $\pm$0.01& 0.70 $\pm$0.01& 0.01\\
               &                &         &         \\
  \multicolumn{4}{c}{${<B>-<V>}$=$C$+$D{[{<V>-<K>}]}$} \\
               &                &         &         \\
0.004 &-0.28 $\pm$0.01& 0.54 $\pm$0.02&  0.01\\
0.008 &-0.24 $\pm$0.01& 0.54 $\pm$0.02& 0.01\\
0.02  &-0.13 $\pm$0.01& 0.52 $\pm$0.01& 0.01\\
\hline
\hline
\end{tabular}
\end{center}
%\end{flushleft}
\end{table}

%\enddata
%\tablenotetext{a}{Metal content.
%\hspace*{0.5mm} $^b$ Zero point.
%\hspace*{0.5mm} $^c$ Second color coefficient.
%\hspace*{0.5mm} $^d$ Standard deviation (mag).
%\hspace*{0.5cm} $^e$ Errors on the coefficients.}
%\end{deluxetable}

%\begin{document}
%\begin{deluxetable}{cccccc}
%\tablehead{
%\colhead{Z\tablenotemark{a}} &
%\colhead{$\alpha$\tablenotemark{b}} &
%\colhead{$\beta$\tablenotemark{c}} &
%\colhead{$\gamma$\tablenotemark{d}} &
%\colhead{$\sigma$\tablenotemark{e}} }
%\startdata

\begin{table}
\caption{Theoretical PLC relations.}
\begin{center}
\begin{tabular}{ccccc}
\hline
\hline
Z & $\alpha$ & $\beta$  & $\gamma$ & $\sigma$\\
\hline
      &         &                &         &         \\
  \multicolumn{5}{c}{$<M_V>$=$\alpha$+$\beta$log$P$+$\gamma$[$<B>-<V>$]} \\
      &         &                &         &         \\
0.004 &-2.54 $\pm$0.04& -3.52 $\pm$0.03& 2.79 $\pm$0.07&  0.04\\
0.008 &-2.63 $\pm$0.04& -3.55 $\pm$0.03& 2.83 $\pm$0.06& 0.03\\
0.02  &-2.98 $\pm$0.07& -3.72 $\pm$0.10& 3.27 $\pm$0.18& 0.07\\
      &          &         &         &         \\
    \multicolumn{5}{c}{$<M_V>$=$\alpha$+$\beta$log$P$+$\gamma$[$<V>-<R>$]} \\
         &         &         &         &         \\
0.004 &-3.28 $\pm$0.03& -3.57 $\pm$0.02& 6.93 $\pm$0.12& 0.03 \\
0.008 &-3.31 $\pm$0.03& -3.59 $\pm$0.02& 6.97 $\pm$0.11& 0.03 \\
0.02  &-3.40 $\pm$0.04& -3.62 $\pm$0.05& 7.09 $\pm$0.18& 0.04 \\
      &          &         &         &         \\
    \multicolumn{5}{c}{$<M_V>$=$\alpha$+$\beta$log$P$+$\gamma$[$<V>-<I>$]} \\
         &         &         &         &         \\
0.004 &-3.55 $\pm$0.03& -3.58 $\pm$0.03& 3.75 $\pm$0.07& 0.03 \\
0.008 &-3.54 $\pm$0.03& -3.59 $\pm$0.02& 3.74 $\pm$0.06& 0.03 \\
0.02  &-3.61 $\pm$0.03& -3.59 $\pm$0.04& 3.85 $\pm$0.09& 0.03 \\
      &          &         &         &         \\
    \multicolumn{5}{c}{$<M_V>$=$\alpha$+$\beta$log$P$+$\gamma$[$<V>-<J>$]} \\
         &         &         &         &         \\
0.004 &-3.47 $\pm$0.03& -3.60 $\pm$0.03& 2.26 $\pm$0.04& 0.03 \\
0.008 &-3.40 $\pm$0.03& -3.60 $\pm$0.02& 2.20 $\pm$0.04& 0.03 \\
0.02  &-3.29 $\pm$0.04& -3.59 $\pm$0.04& 2.11 $\pm$0.05& 0.03 \\
      &          &         &         &   \\
    \multicolumn{5}{c}{$<M_V>$=$\alpha$+$\beta$log$P$+$\gamma$[$<V>-<K>$]} \\
         &         &         &         &         \\
0.004 &-3.44 $\pm$0.04& -3.61 $\pm$0.03& 1.64 $\pm$0.03& 0.04\\
0.008 &-3.37 $\pm$0.04& -3.60 $\pm$0.03& 1.61 $\pm$0.03& 0.03 \\
0.02  &-3.25 $\pm$0.04& -3.55 $\pm$0.05& 1.53 $\pm$0.04& 0.04\\
\hline
\hline
\end{tabular}
\end{center}
%\end{flushleft}
\end{table}

\begin{table}
\caption{Theoretical Wesenheit relations.}
\begin{center}
\begin{tabular}{cccc}
\hline
\hline
Z & $a$ & $b$  & $\sigma$\\
\hline
               &                &         &         \\
  \multicolumn{4}{c}{$W(B,V)$=$<M_V>$-3.1[$<B>-<V>$]=$a$+$b$log$P$} \\
               &                &         &         \\
0.004 &-2.61 $\pm$0.05& -3.65 $\pm$0.02&  0.05\\
0.008 &-2.71 $\pm$0.04& -3.67 $\pm$0.02& 0.04\\
0.02  &-2.93 $\pm$0.07& -3.63 $\pm$0.03& 0.07\\
                &         &         &         \\
  \multicolumn{4}{c}{$W(V,R)$=$<M_R>$-5.29[$<V>-<R>$]=$a$+$b$log$P$} \\
               &                &         &         \\
0.004 &-3.15 $\pm$0.04& -3.46 $\pm$0.02&  0.04\\
0.008 &-3.17 $\pm$0.04& -3.46 $\pm$0.02& 0.04\\
0.02  &-3.24 $\pm$0.05& -3.43 $\pm$0.02& 0.05\\
                &         &         &         \\
  \multicolumn{4}{c}{$W(V,I)$=$<M_I>$-1.54[$<V>-<I>$]=$a$+$b$log$P$} \\
               &                &         &         \\
0.004 &-3.00 $\pm$0.10& -3.21 $\pm$0.05&  0.10\\
0.008 &-2.99 $\pm$0.10& -3.17 $\pm$0.04& 0.10\\
0.02  &-3.04 $\pm$0.10& -3.02 $\pm$0.04& 0.10\\
                &         &         &         \\
  \multicolumn{4}{c}{$W(V,J)$=$<M_J>$-0.39[$<V>-<J>$]=$a$+$b$log$P$} \\
               &                &         &         \\
0.004 &-2.86 $\pm$0.12& -3.15 $\pm$0.05&  0.12\\
0.008 &-2.83 $\pm$0.11& -3.12 $\pm$0.04& 0.11\\
0.02  &-2.83 $\pm$0.10& -3.02 $\pm$0.04& 0.10\\
                &         &         &         \\
  \multicolumn{4}{c}{$W(V,K)$=$<M_K>$-0.13[$<V>-<K>$]=$a$+$b$log$P$} \\
               &                &         &         \\
0.004 &-2.95 $\pm$0.10& -3.24 $\pm$0.05&  0.10\\
0.008 &-2.92 $\pm$0.09& -3.21 $\pm$0.04& 0.09\\
0.02  &-2.90 $\pm$0.08& -3.12 $\pm$0.03& 0.08\\
\hline
\hline
\end{tabular}
\end{center}
%\end{flushleft}
\end{table}

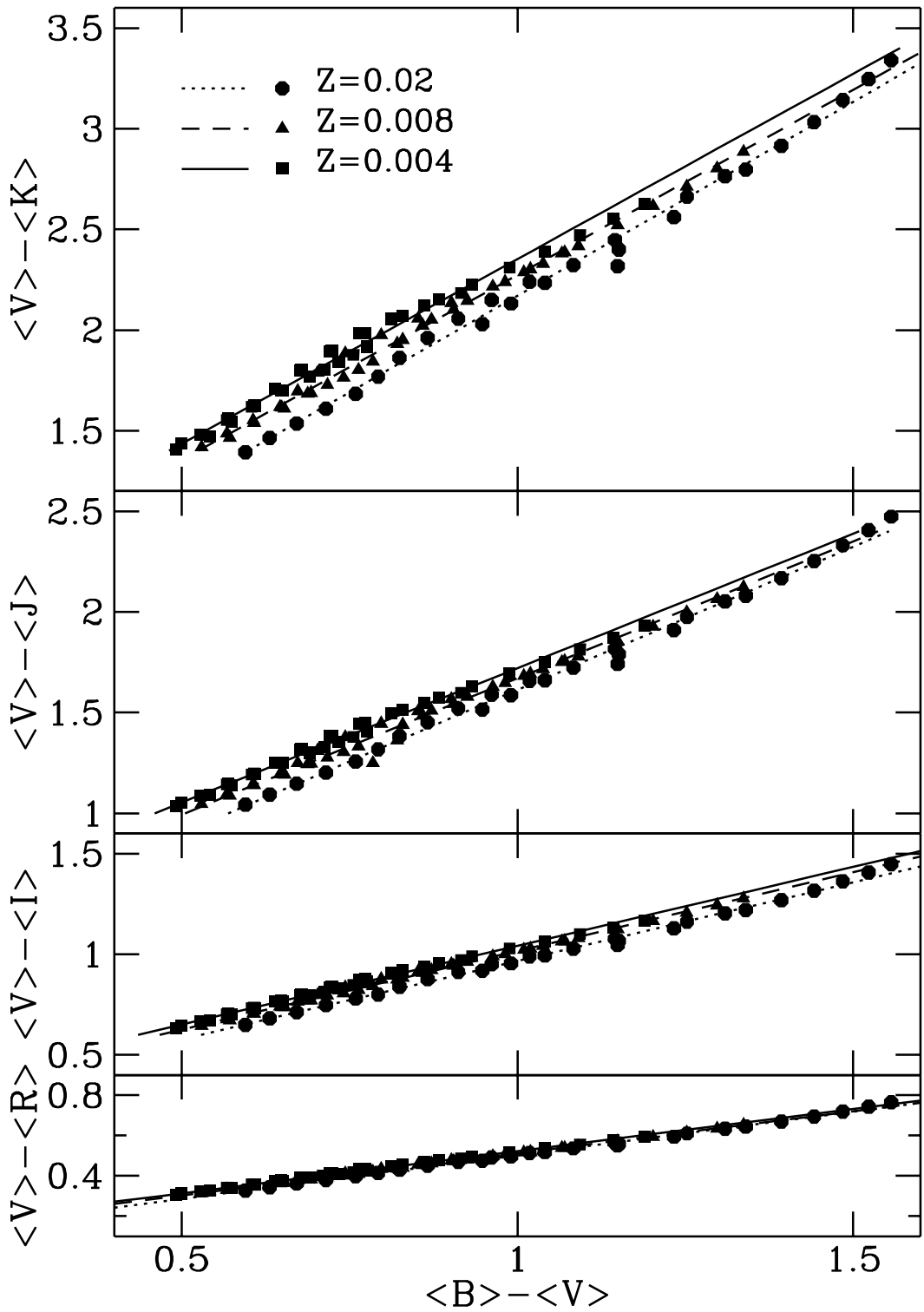
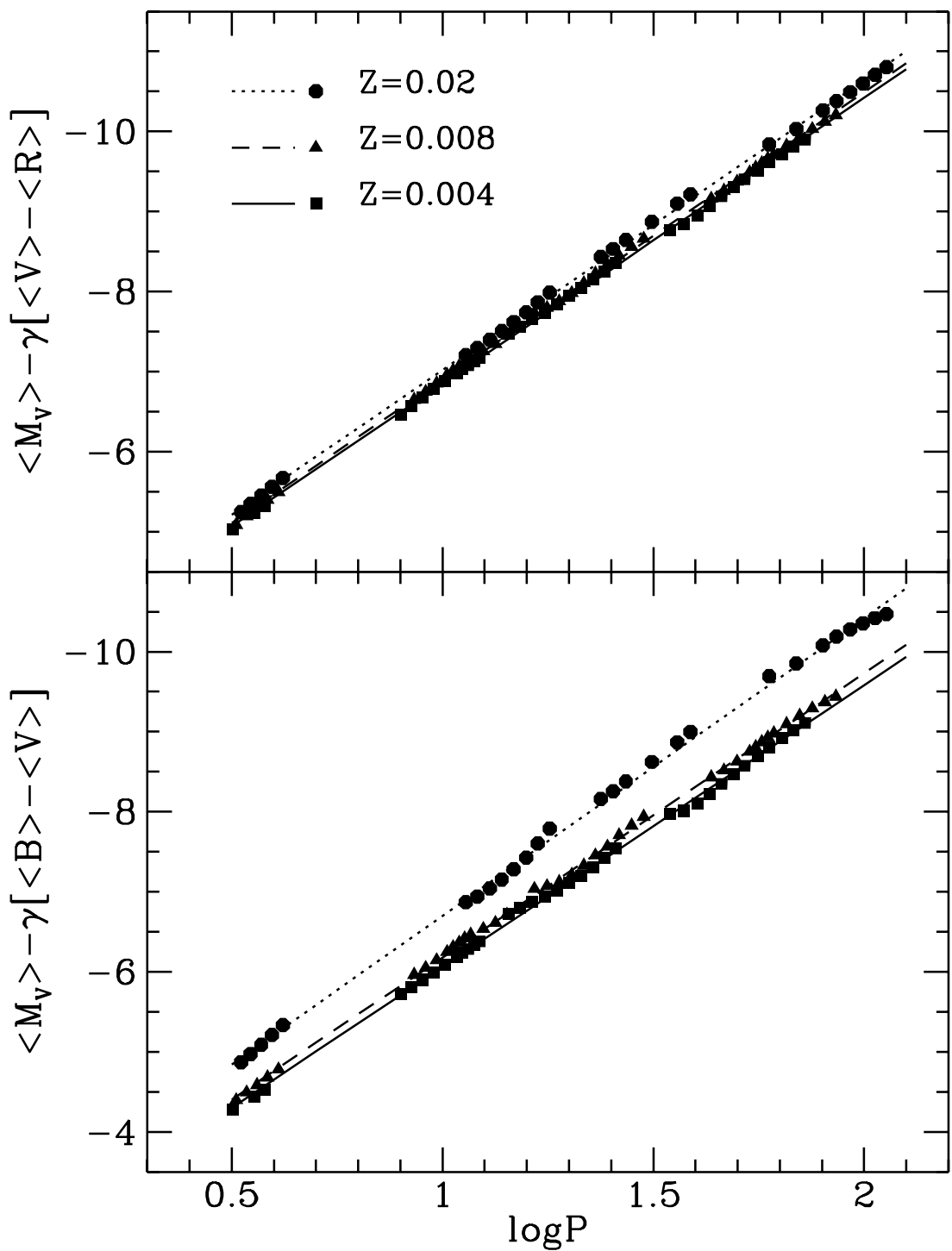
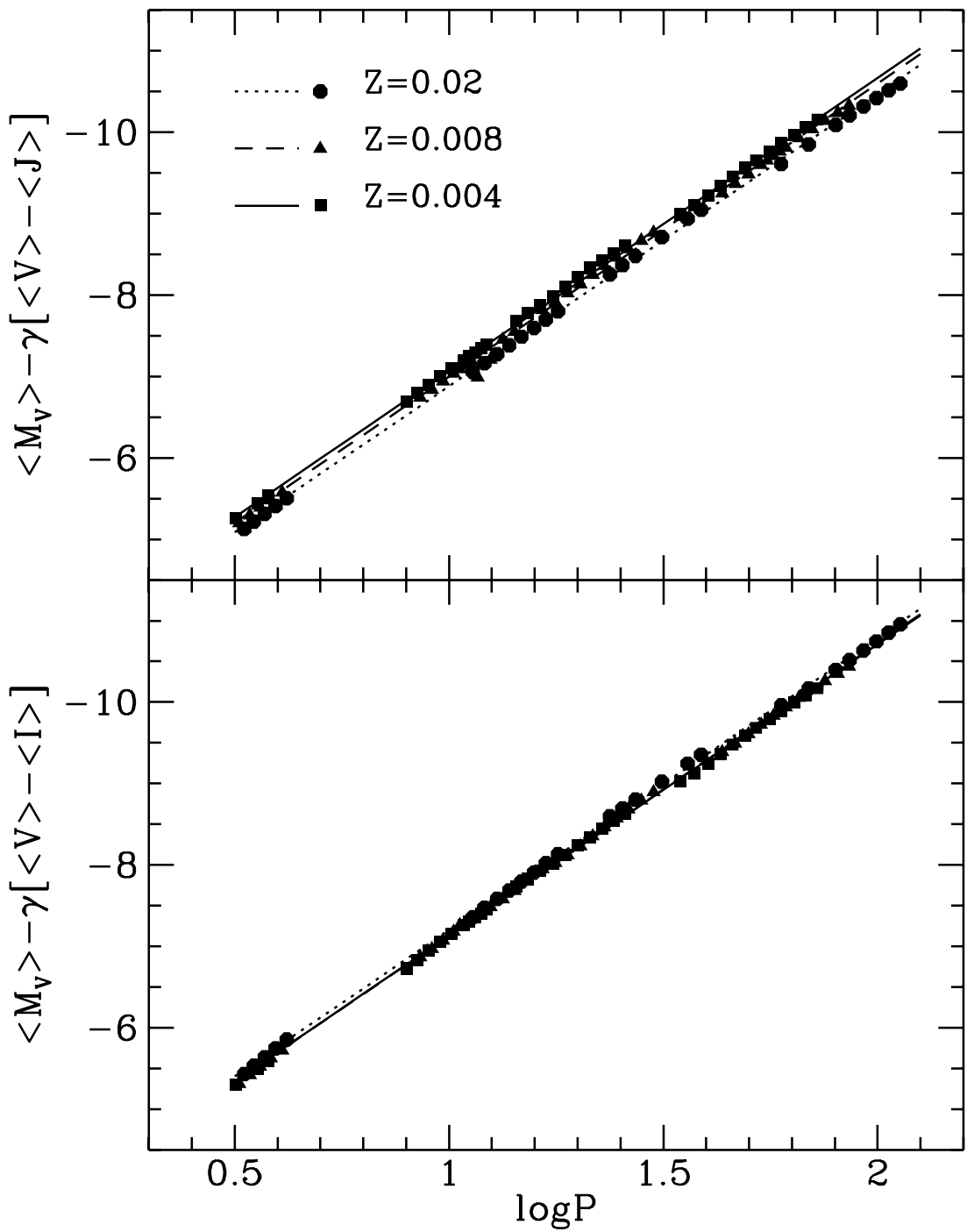
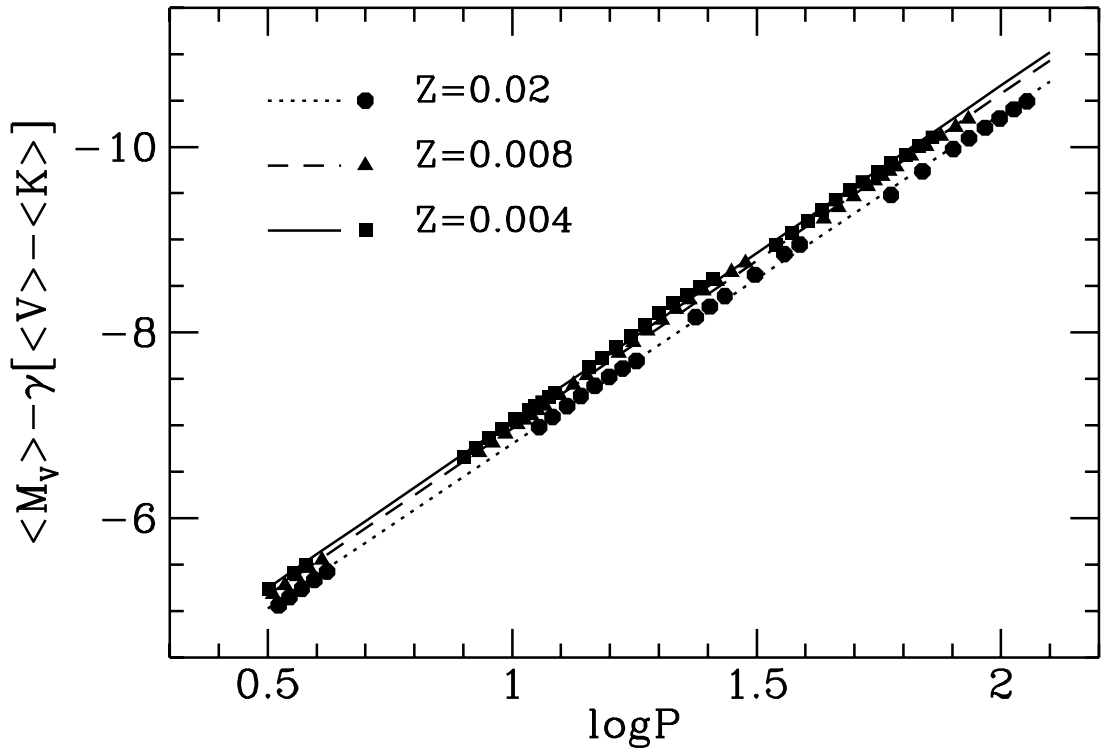
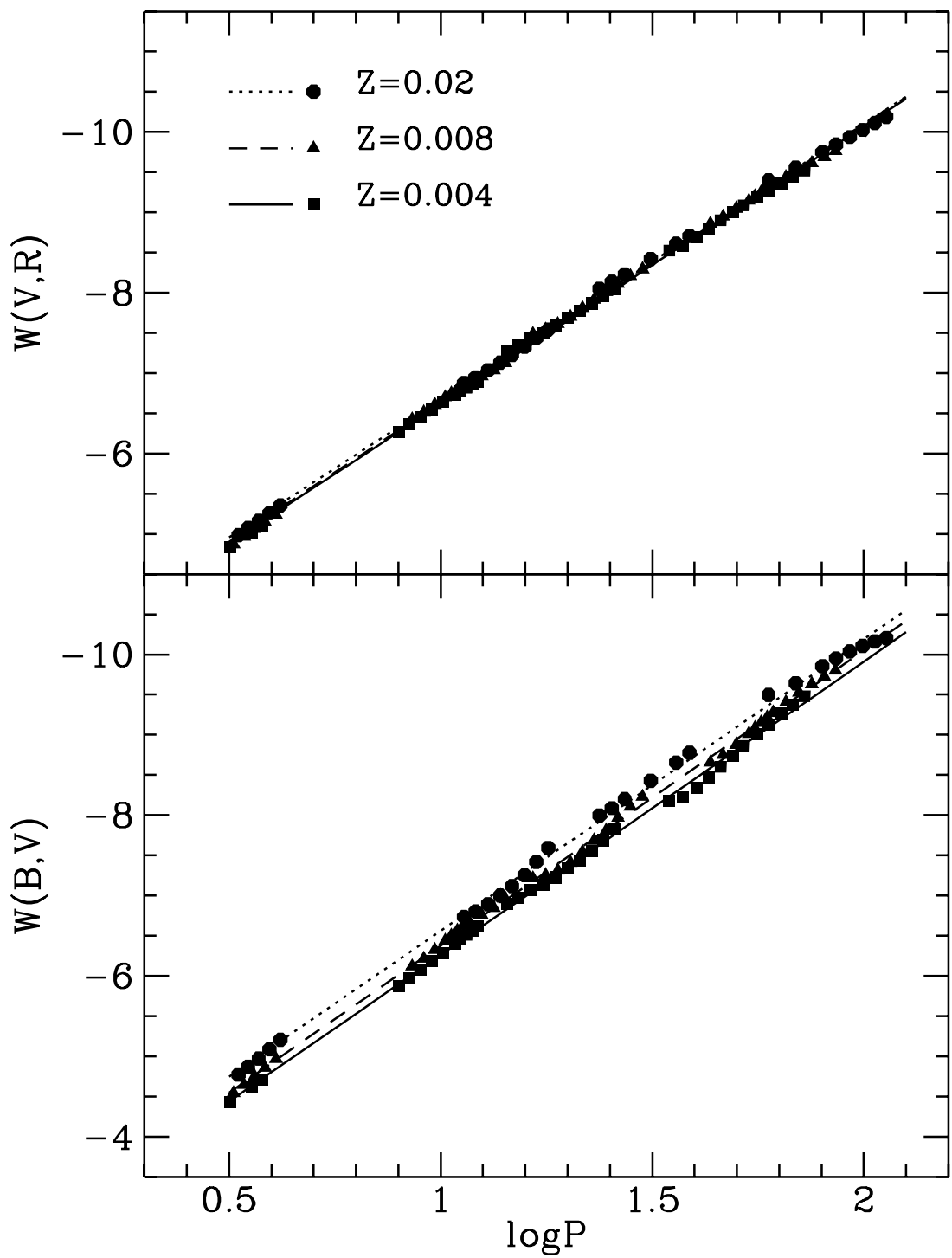
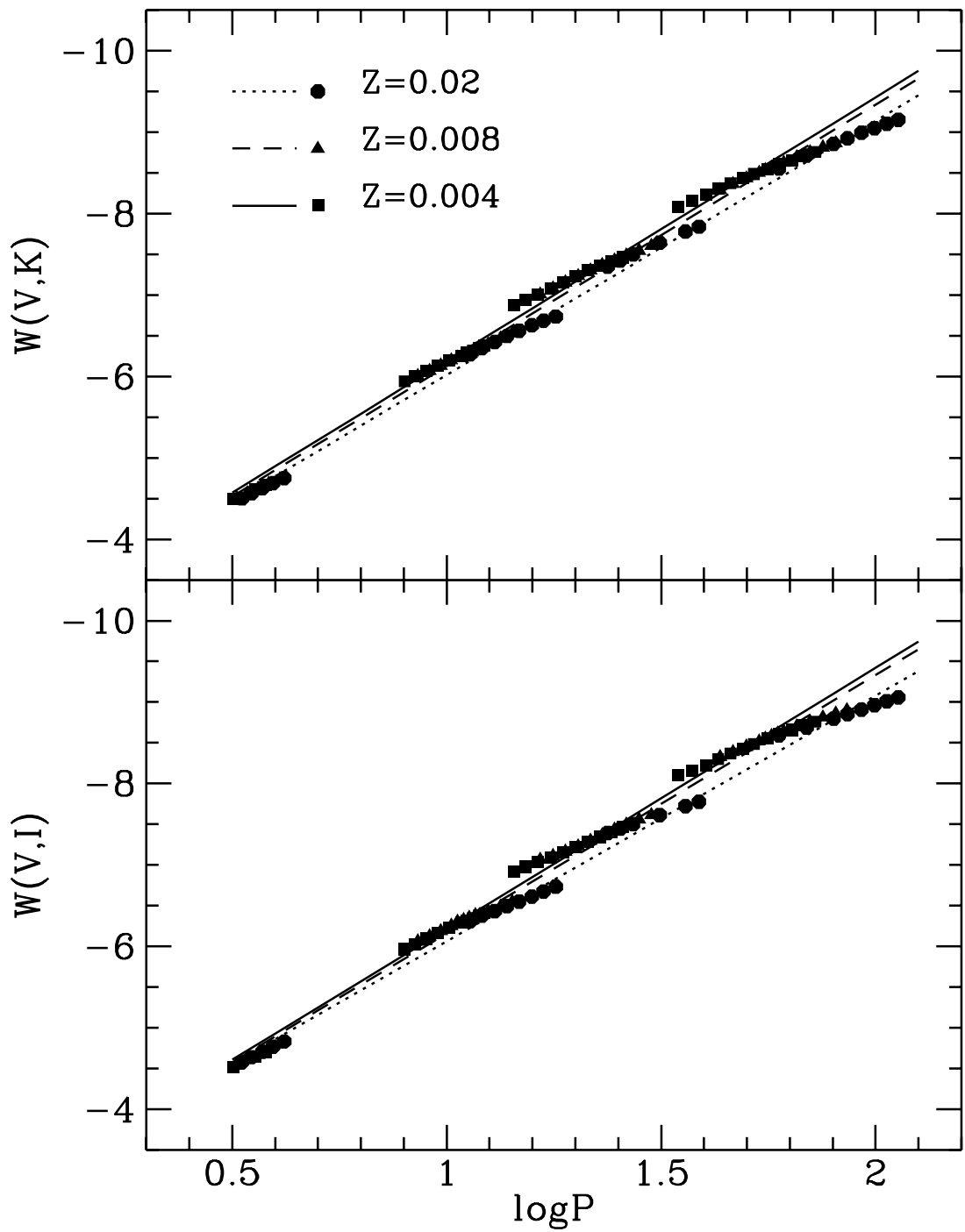
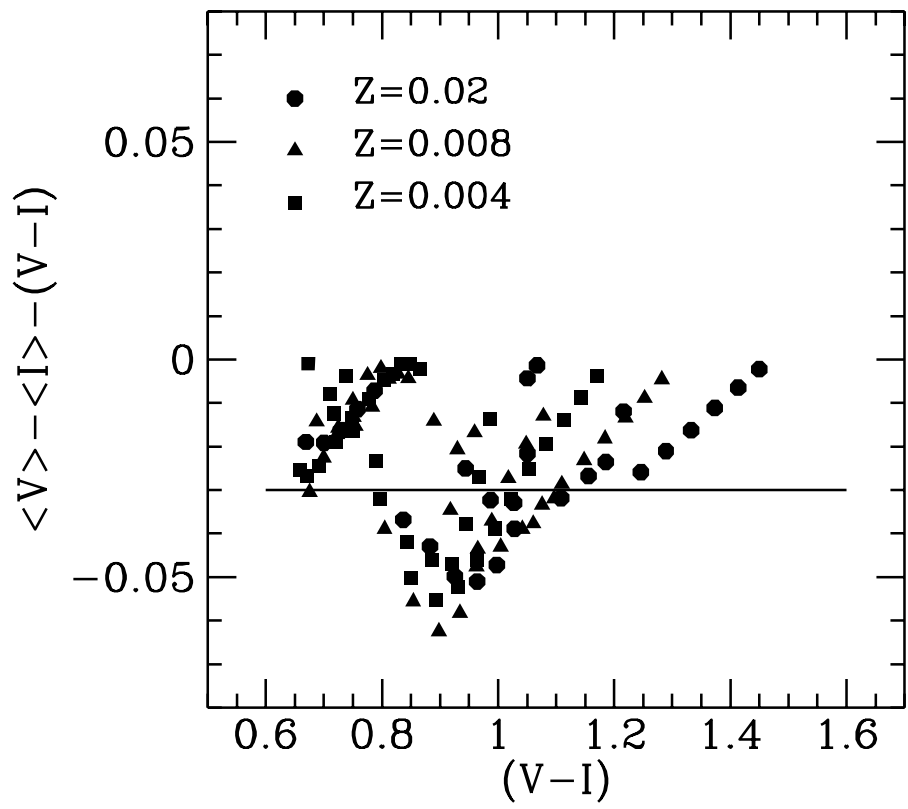
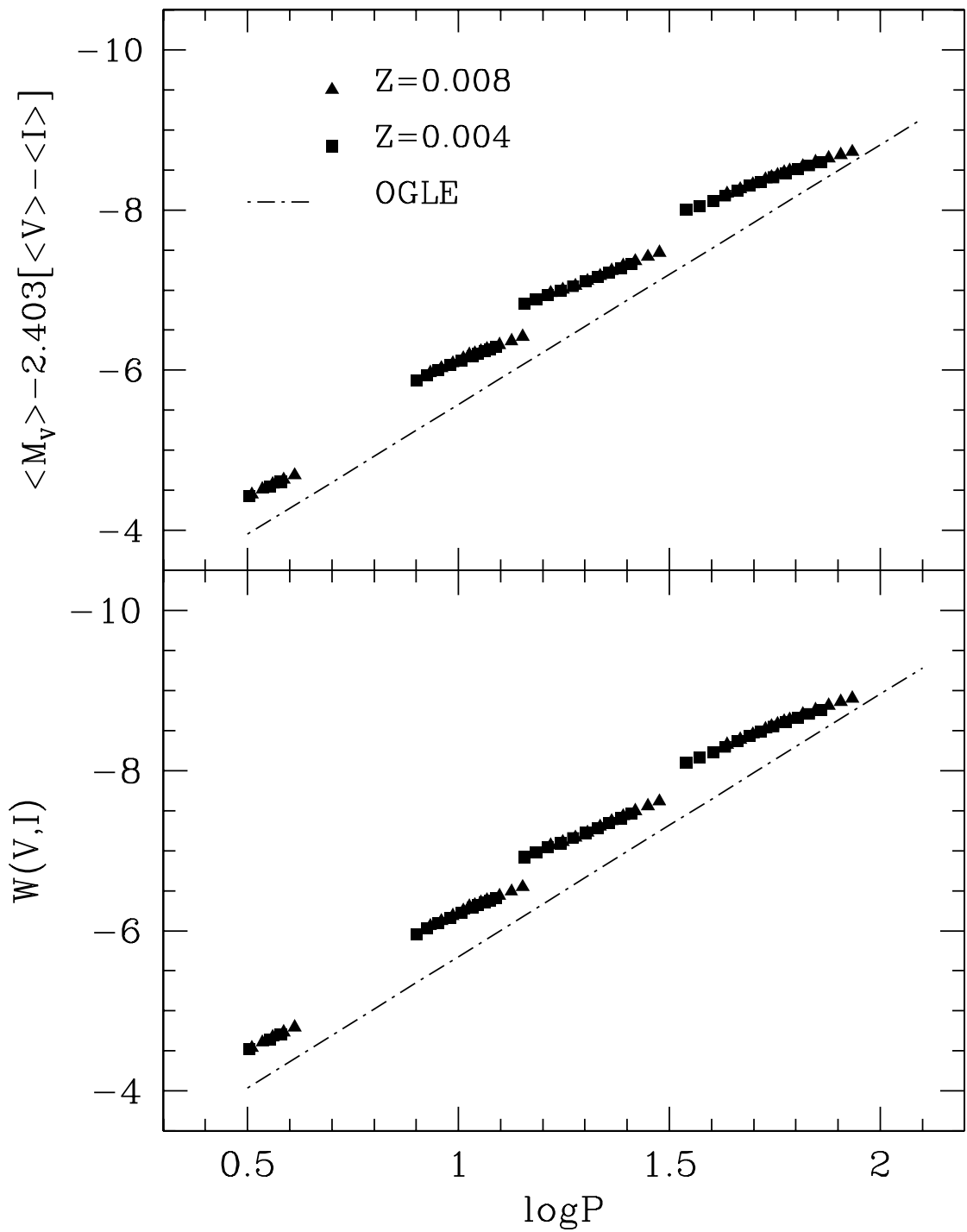
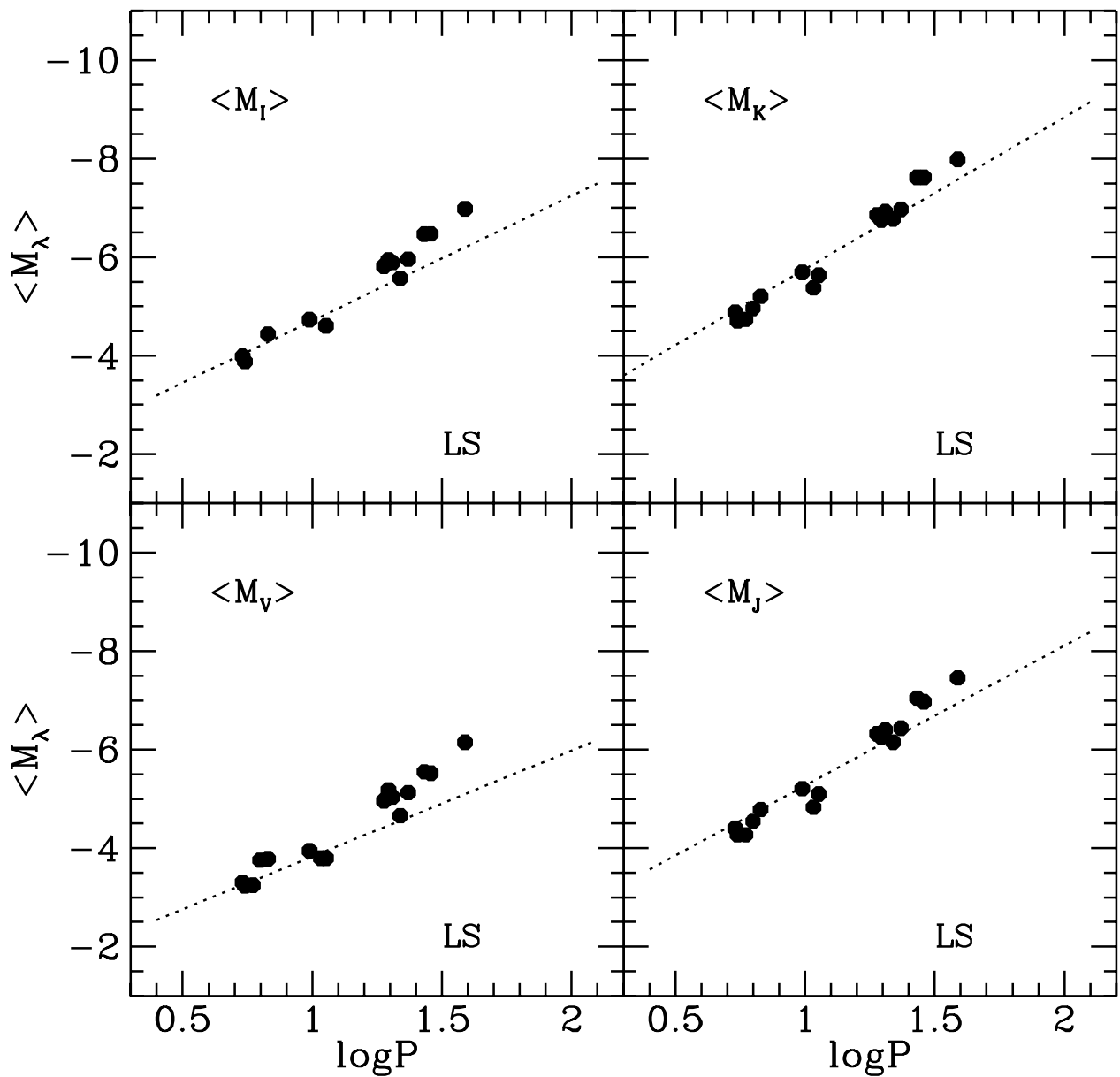
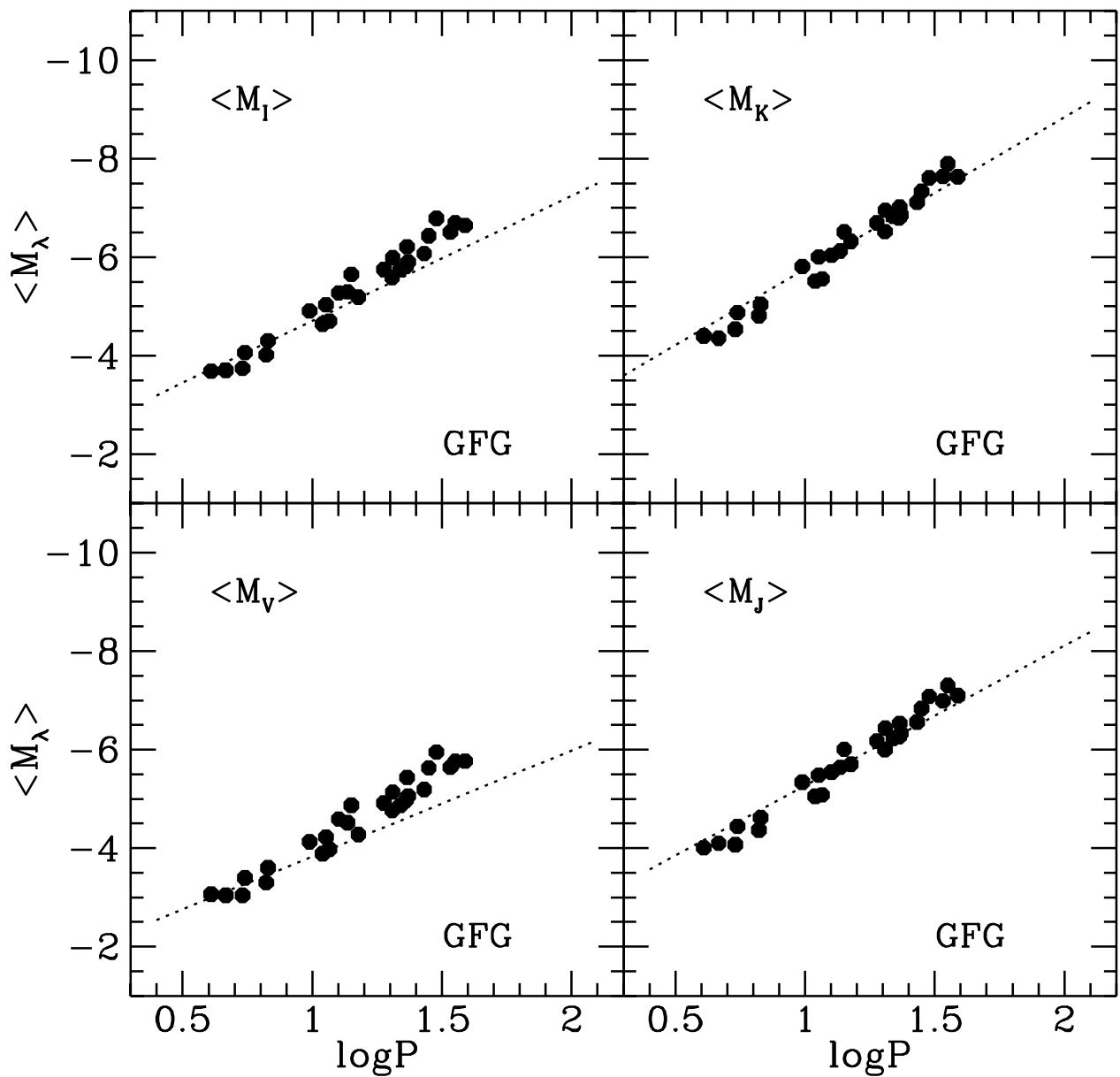
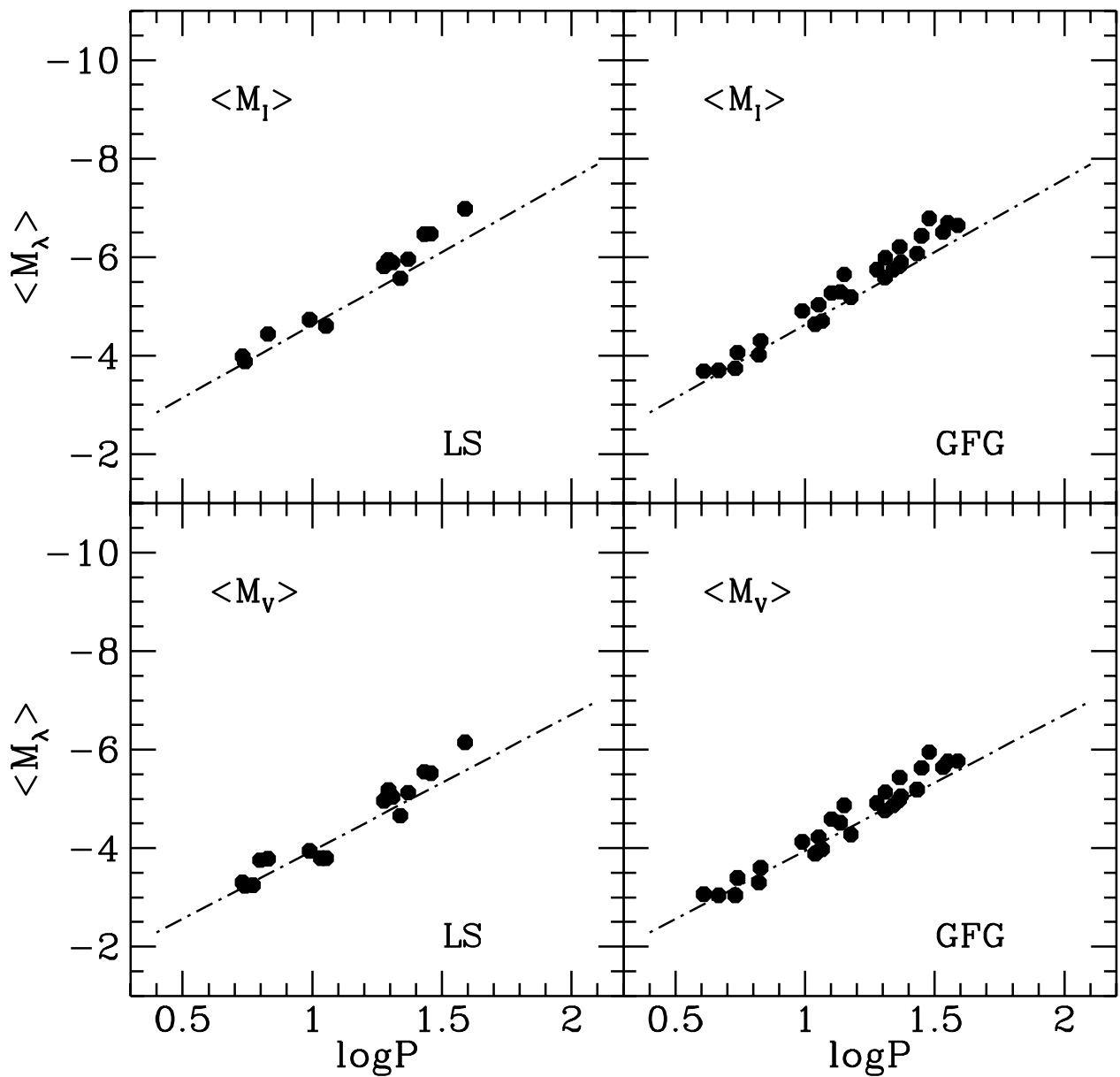
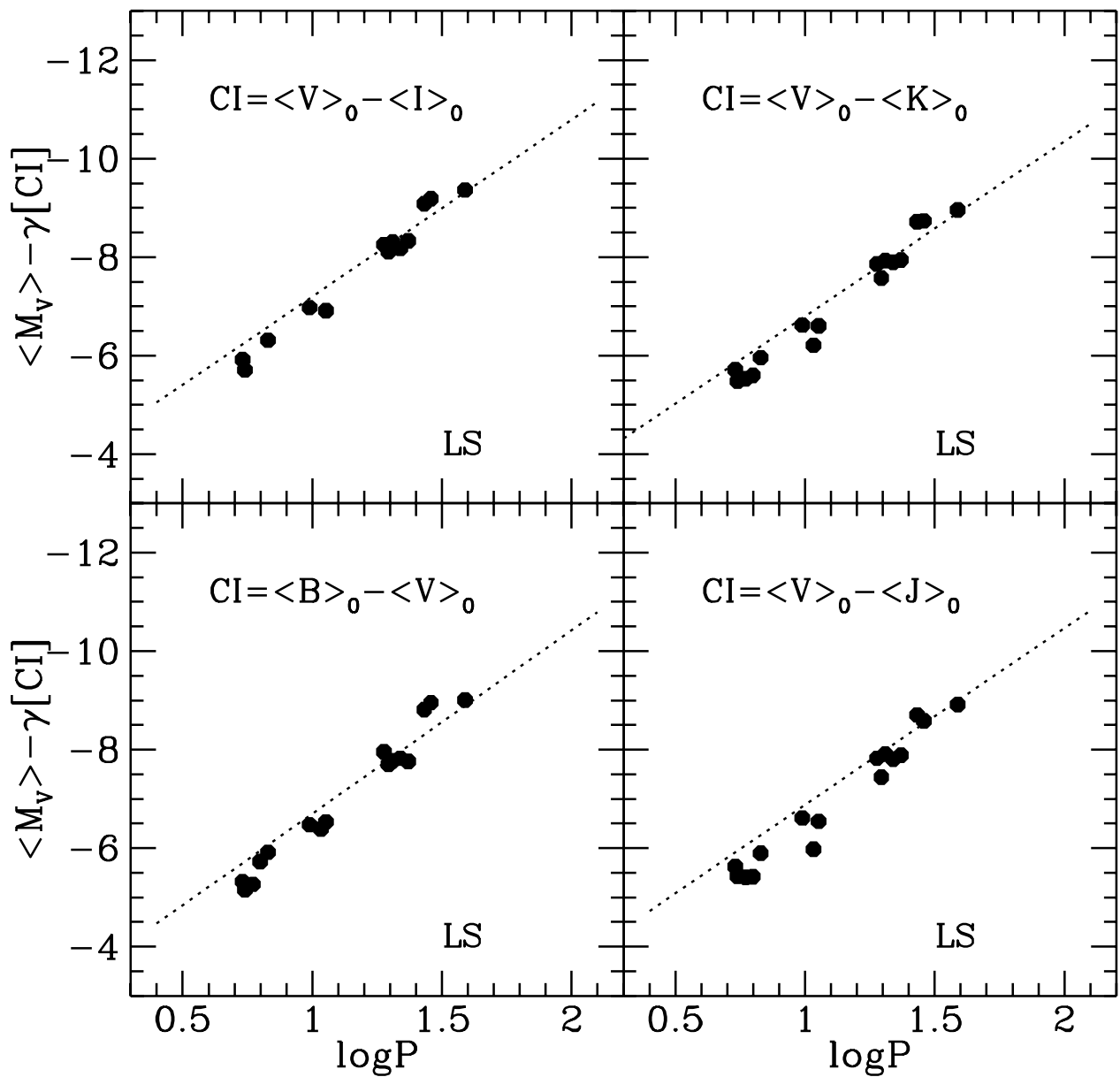
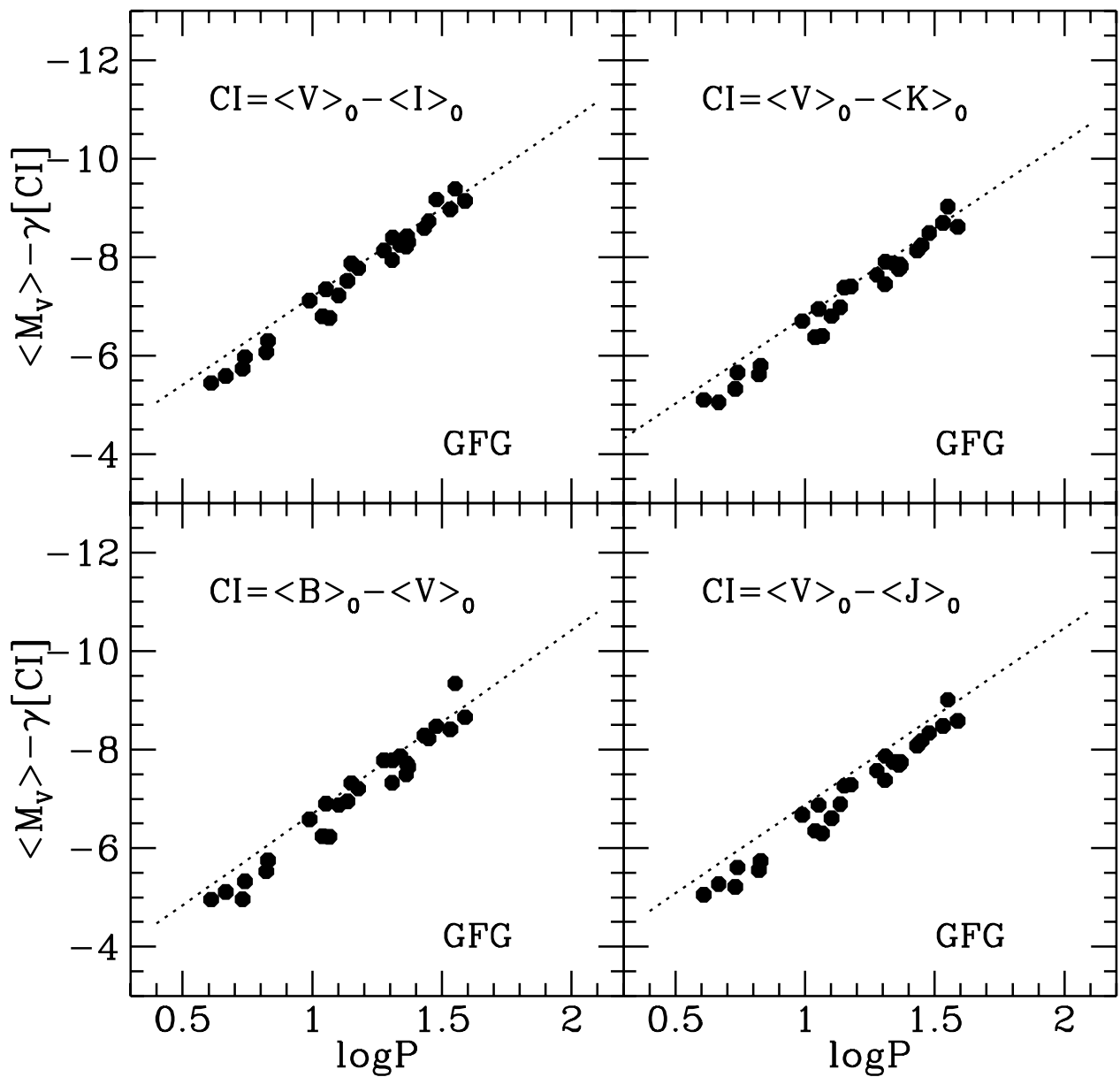
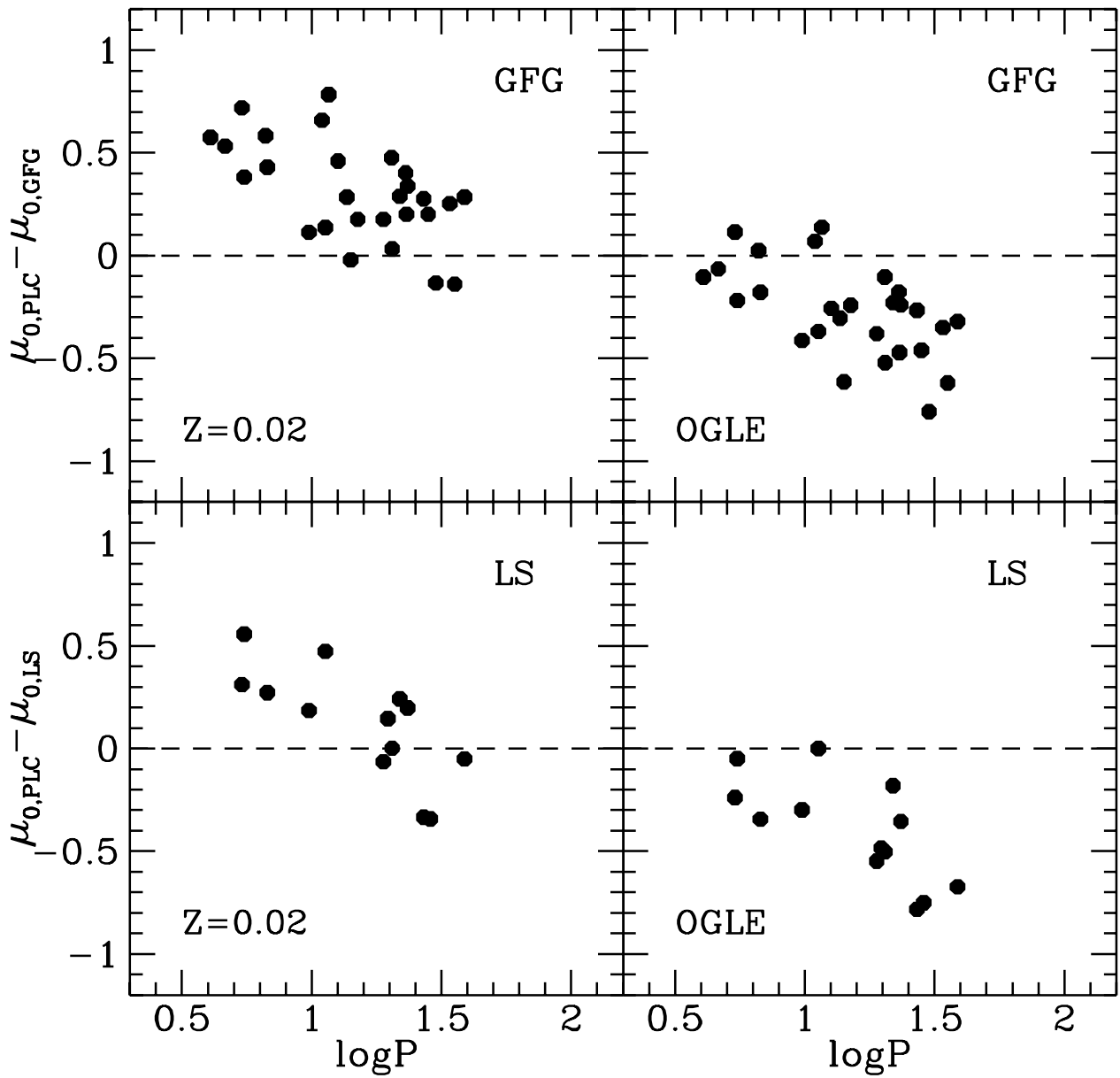
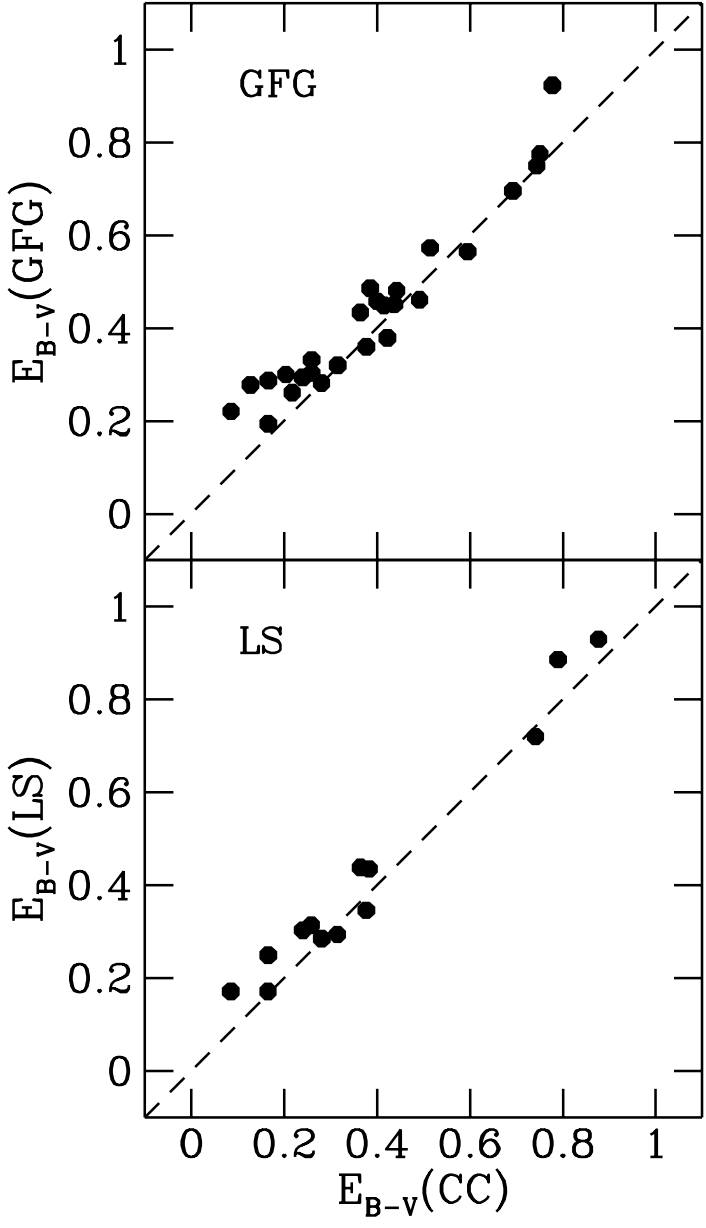
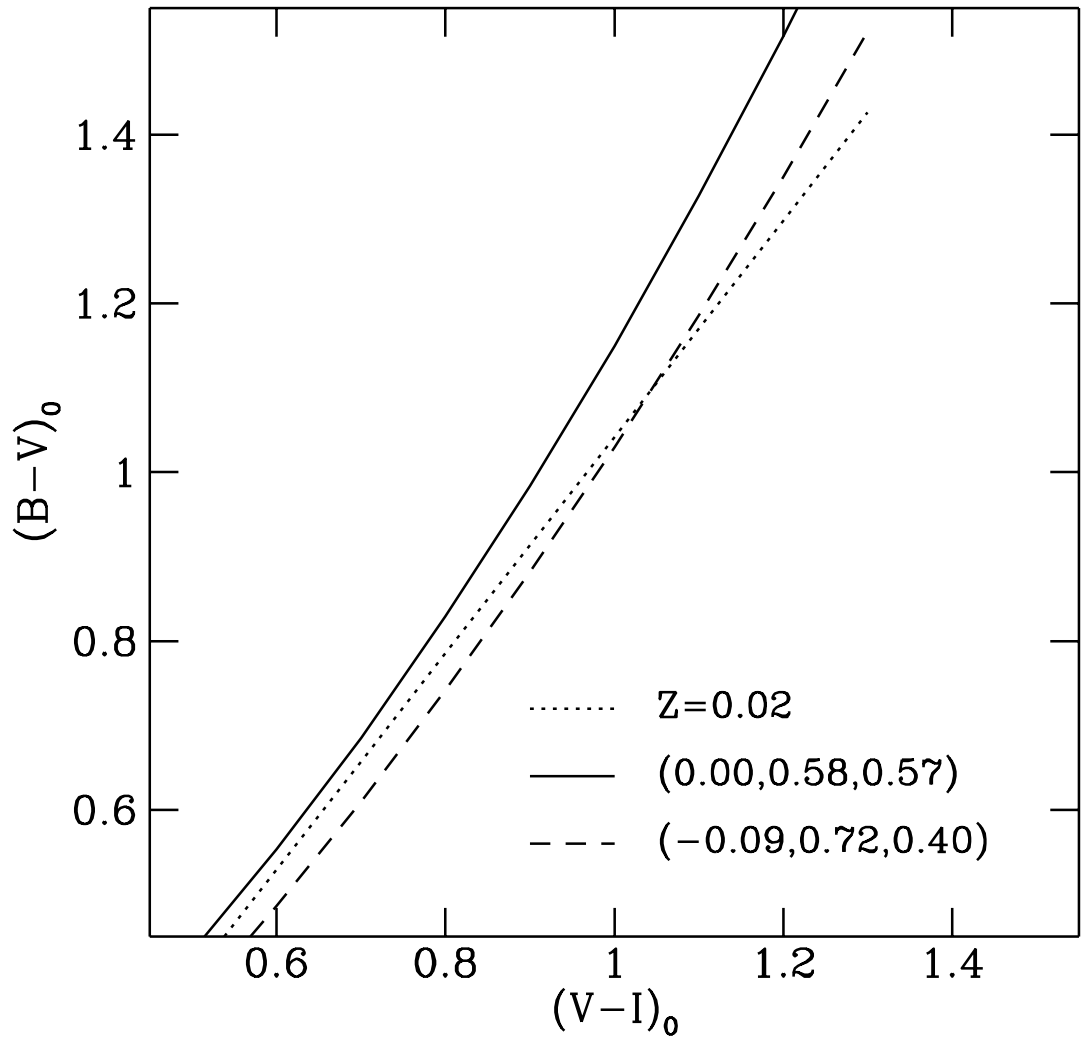
\end{document}